\documentclass[10pt]{article}

\usepackage{amsmath}
\usepackage{amssymb}

\usepackage{graphicx}

\usepackage{cite}

\usepackage{color} 

\usepackage[labelfont=bf,labelsep=period,justification=raggedright]{caption}

\makeatletter
\renewcommand{\@biblabel}[1]{\quad#1.}
\makeatother

\date{}

\usepackage{graphicx}
\usepackage{dcolumn}
\usepackage{bm}
\usepackage{amssymb}
\usepackage{amsfonts}

\begin{document}

\begin{flushleft}
{\Large
\textbf{From gene regulatory networks to population dynamics: robustness, diversity and their role in progression to cancer}
}
\\
\mbox{}\\
\bf Tom\'as Alarc\'on$^{1,\ast}$ and Henrik Jeldtoft Jensen$^{2}$
\\\mbox{}\\
\bf{1} Basque Centre for Applied Mathematics, Bizkaia Technology Park, 48160 Derio, Bizkaia, Spain
\\
\bf{2} Institute for Mathematical Sciences, Imperial College London, 53 Princes' Gate, London SW7 2PG, United Kingdom
\\
$\ast$ E-mail: Corresponding author alarcon@bcamath.org
\end{flushleft}

\section*{Abstract}
The aim of this paper is to discuss the role of robustness and diversity in population dynamics in particular to some properties of the multi-step from healthy tissue to fully malignant tumours. Recent evidence shows that diversity within the cell population of a neoplasm, a pre-tumoural lession that can develop into a fully malignant tumour, is the best predictor for its evolving into a tumour. By studying the dynamics of a population described by a multi-type, population-size limited branching process in terms of the evolutionary formalism, we show some general principles regarding the probability of a resident population to being invaded by a mutant population in terms of the number of types present in the population and their resilience. We show that, although diversity in the mutant population poses a barrier for the emergence of the initial (benign) lession, under appropiate conditions, namely, the phenotypes in the mutant population being more resilient than those of the resident population, a more variable noeplastic population is more likely to be invaded by a more malignant one. Analysis of a model of gene regulatory networks suggest possible mechanisms giving rise to mutants with increased phenotypic diversity and robustness. We then go on to show how these results may help us to interpret some recent data regarding the evolution of Barrett's oesophagus into throat cancer.
\section*{Author Summary}
Recent results by Maley and co-workers \cite{maley2006} regarding progression of the Barrett's esophagus, a benign, neoplastic throat condition, into a malignant tumour have revealed that clonal diversity within diseased tissue is the best risk predictor outperforming usual genetic markers. So far, there is no detailed explanation as to why this is so. What is the role of diversity in the progression to cancer and what are the mechanisms involved? In this paper, we address these issues and put forward some generic mechanisms that may help to gain a better understanding of the empirical results obtained by Maley et al. Our approach to this problem is in two steps. First, we propose a simple model of gene regulatory networks and analyse their behaviour upon gene silencing. We observe that, following gene silencing, both phenotypic diversity and robustness increase. We then proceed to analyse how these two effects affect the dynamics of the cell populations and find that more diverse pre-malignant lessions are likely to evolve into a more malignant form only if phenotypic robustness increases, hence suggesting that robustness is essential to the results put forward by Maley and coworkers. 
\section*{Introduction}
Complex diseases such as cancer pose an enormous scientific challenge. Cancer involves an extra level of complexity as its development involves evolutionary processes driven by Darwinian natural selection \cite{nowak2006}. The traditional view of the emergence of cancer consists of a series of mutations with each of them increasing the fitness of the mutated cells with respect to that of their normal counterparts, which are eventually taken over by succesive rounds of clonal expansion \cite{attolini2009,shibata2006}. Recent experimental analysis on the genetic landscape of tumours \cite{wood2007}, however, show an even more complex situation whereby tumours have been found to be much more genetically diverse than expected where low-frequency mutations are likely to play a leading part in the evolution of tumours. Furthermore, recent analysis of data from patients of Barrett's oesophagus, a neoplastic condition that may evolve into throat cancer, has found that the best predictor for its evolution into a fully malignant lession is diversity in the cell population, outscoring all of the usual genetic markers \cite{maley2006}. In particular, Maley et al. have shown that quantities normally used in ecology to quantify diversity, such as the Shannon entropy, $H_S$, and the number of clones detected in tissue samples extracted from patients of Barrett's oesophagus, are excellent predictors. According to \cite{shibata2006}, these findings bring into question the vision of tumour evolution as a series of succesive rounds of mutation/clonal evolution to favour a model in which premalignant lessions sustain a number of co-existing cellular types.

In order to explore some of the issues put forward by results of Maley et al. \cite{maley2006}, in particular their results concerning the higher probability of pre-malignant lessions with larger diversity to evolve into tumours, we first analyse the robustness process of a simple model of gene regulatory networks. We focus on the study of the effects of gene silencing on both the number of stable phenotypes and their robustness \cite{rutherford1998,siegal2002,bergman2003,ciliberti2007,wagner2008a,levy2008}. We will show using a simple model, similar to the ones used in previous related works \cite{wagner1996,siegal2002,bergman2003,ciliberti2007}, that such modification on the gene regulatory network leads to an increase on both the number of stable (robust) phenotypes and their ability to sustain further mutations (robustness).

We then analyse how these changes induced by gene silencing have on the dynamics of the corresponding population, in particular on the ability of one mutant carrying such modifications to invade the resident cell population. To so doing, we consider a model of population dynamics consisting of a multi-type branching process with population-size limited proliferation probability \cite{alarcon2009}. Our aim is to put forward a number of general properties of multi-type populations which could shed some light on the mechanisms involved in the transition from neoplasm to malignant tumours. Our analysis is carried out in terms of the so-called evolutionary formalim developed by Demetrius and coworkers \cite{arnold1994,demetrius2004,demetrius2009}. This framework extends the thermodynamic formalism developed within ergodic theory and statistical physics \cite{ruelle1978} and it allows us to study the evolutionary behaviour of structured populations in terms of the \emph{evolutionary entropy}, which is a measure of the \emph{dynamical} diversity of the population. 

The evolutionary formalism, whose main results concerning the present problem are summarised in Materials \& Methods, has been recently applied to analyse the competition between resident and mutant populations modelled in terms of age-structured population models \cite{demetrius2009} and multi-type branching processes \cite{alarcon2009}. Demetrius et al. \cite{demetrius2004,demetrius2009} have shown that the probability of a mutant structured population to invade the resident one is determined by the rate at which a population returns to its corresponding steady state after a random perturbation of the parameters that characterise the population dynamics. In turn, it has been proofed that such rate is given by the evolutionary entropy \cite{demetrius2004}. They have also shown that the evolutionary stability of a population, that is, its resilience to invasion by a mutant, is determined by the evolutionary entropy \cite{demetrius2009}. 

Our aim in this paper is to understand the dynamical mechanisms involved in the findings of Maley et al. \cite{maley2006}. In other words, we intend to find how the Shannon entropy, which is an index of diversity within the population, relates to the evolutionary entropy, which determines the evolutionary stability (or lack thereby) of the corresponding population.  
\section*{Results}

\subsection*{Robustness of gene regulatory networks}
In order to model the dynamics of gene regulatory models, we consider a network with $N_G$ nodes, each representing a gene. To each of the nodes we associate a state $g_i=\pm 1,i=1,\dots,N_G$, with state $g=1(-1)$ corresponding to the gene being activated (inactivated). We generate a directed network, defined in terms of the corresponding adjacency matrix, according to a prescribed degree distribution (in the present case the in- and out-degrees are exponentially distributed). The links of this network correspond to the interactions between genes, which can be activatory or inhibitory. If a gene $i$ activates gene $j$, the corresponding link, represented in a matrix ${\sf W}=(w_{ij})$ whose entries are zero if there is no link between nodes $i$ and $j$ and non-zero otherwise, carries a weight $w_{ij}=1$. If, on the contrary, gene $i$ inhibits gene $j$, the corresponding weight $w_{ij}=-1$. In our model, positive and negative weights are randomly assigned with probability $p_+$ and $1-p_+$, respectively. 

The dynamics of the gene regulatory netowrk is defined in terms of the above elements as follows. An initial condition $g_i(t=0),i=1,\dots,N_G$ and a matrix ${\sf W}$ are given. For each node and at each time step the following quantity is computed: $I_i(t)=\sum_{j\in\langle i \rangle_{in}}w_{ji}g_j(t)$. Then at each time step:

\begin{equation}\label{eq:gnr1}
g_i(t+1)=\left\{\begin{array}{l}1\mbox{ if }I_i(t)>0\\-1\mbox{ if }I_i(t)<0\\g_i(t)\mbox{ if }I_i(t)=0\end{array}\right.
\end{equation}
 
\noindent which is ran until the system reaches a steady state. The phenotype, $\phi$, is defined as the vector $\phi=(g_1,\dots,g_{N_G})$ at steady state. This steady state may be a periodic solution. How we deal with this type of solution will be explained in more detail later.

In our model, we consider two type of mutations. On the one hand, following \cite{wagner1996,siegal2002,ciliberti2007}, we introduce random changes on how genes regulate each other (i.e. $w_{ij}\rightarrow -w_{ij}$ with a certain probability). The second type of mutation we consider is a form of gene silencing where one particular gene is chosen (in principle, at random) and its state is fixed at $g_i=-1$ regardless of the state what remains of the network. This event could model both loss-of-function mutations \cite{nowak2006} or an epigenetic silencing event \cite{jablonka2005}. 

\begin{figure}
\begin{center}
\includegraphics[scale=0.5]{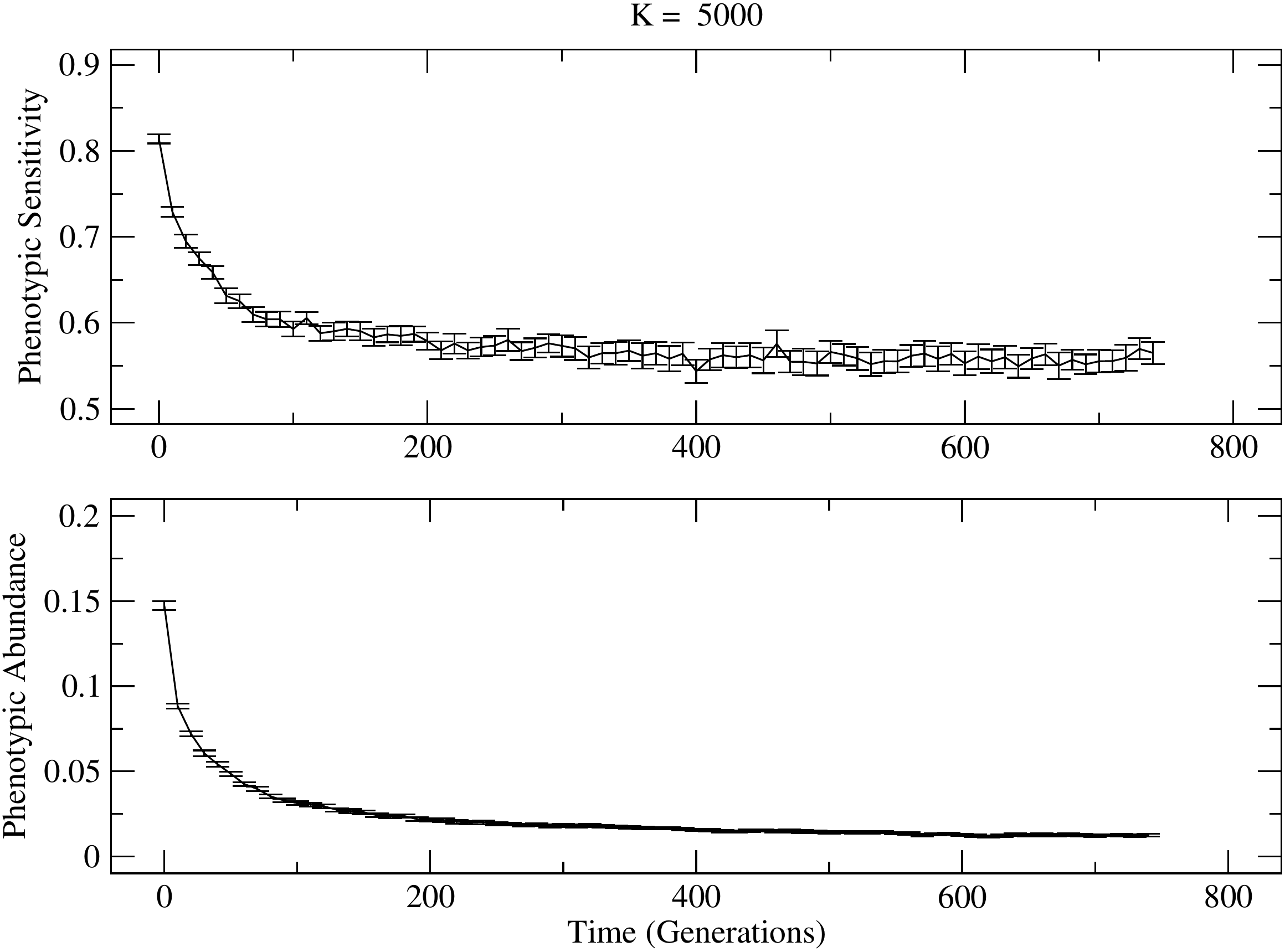}
\end{center}
\caption{Simulations of the evolution of robustness and phenotypic diversity in a model of gene regulatory networks (see text for details). The upper plot corresponds to the phenotypic sensitivity to mutations, i.e. the probability that a mutation induces a change in phenotype, which is an inverse measure of robustness: the more sensitive, the less robust the phenotype. The lower plot corresponds to the phenotypic abundance defined as the number of phenotypes that are actually present within the population at generation $t$ divided by the total population. We observe that both quantities decrease with time. Parameter values: $N_G=10$, $\mu=K^{-1}=0.0002$, $r_m=1$ \label{fig:gnr-nosil}}
\end{figure}

In order to analyse the robustness properties of the corresponding phenotypes, $\phi$, which we define as the probability that a mutation of the first type according to the description in the paragraph above induces a phenotype switch (defined as a change in sign of at least one of the components of the vector $\phi$), we carry out a numerical experiment using the algorithm defined in the Materials \& Methods section. Note that the population dynamics defined by this algorithm is completely neutral, since all the individuals yield offspring with the same probability and we have not imposed any viability conditions as assume e.g. in \cite{siegal2002}. The corresponding results are shown in Fig. \ref{fig:gnr-nosil}. We see that even in neutral conditions the diversity of phenotypes within the population decreases, while the extant types exhibit a greater resilience  against mutations. In other words, the number of different phenotypes decreases with time but the surviving ones are such that their robustness increases, i.e. we observe the emergence of canalisation \cite{waddington1942,jablonka2005} within our simple model. 

These results can be understood in terms of the concept of the neutral network as introduced in \cite{ciliberti2007b,wagner2007}. The neutral network is a meta-network where each node corresponds to a different choice of ${\sf W}$ and two networks are connected if they differ only in one of the values of the corresponding entries $w_{ij}$. It has been shown \cite{ciliberti2007b} that phenotypic robustness can be understood in terms of the structure of such network: robust phenotypes are those whose subnetwork is highly interconnected, i.e. mutation events are likely to produce a network with the same phenotype. This means that robust phenotypes act like traps within this network making it harder for mutations to drive the system out of the corresponding phenotype. The same topological mechanism causes  non-robust phenotypes to disappear at early stages in the evolution of the population. This is because mutations will drive them towards  network configurations representing the more stable phenotypes. Thus, we can qualitatively explain both why the number of phenotypes decreases in time and also how the remaining phenotypes become more resilient.

\begin{figure}
\begin{center}
\includegraphics[scale=0.5]{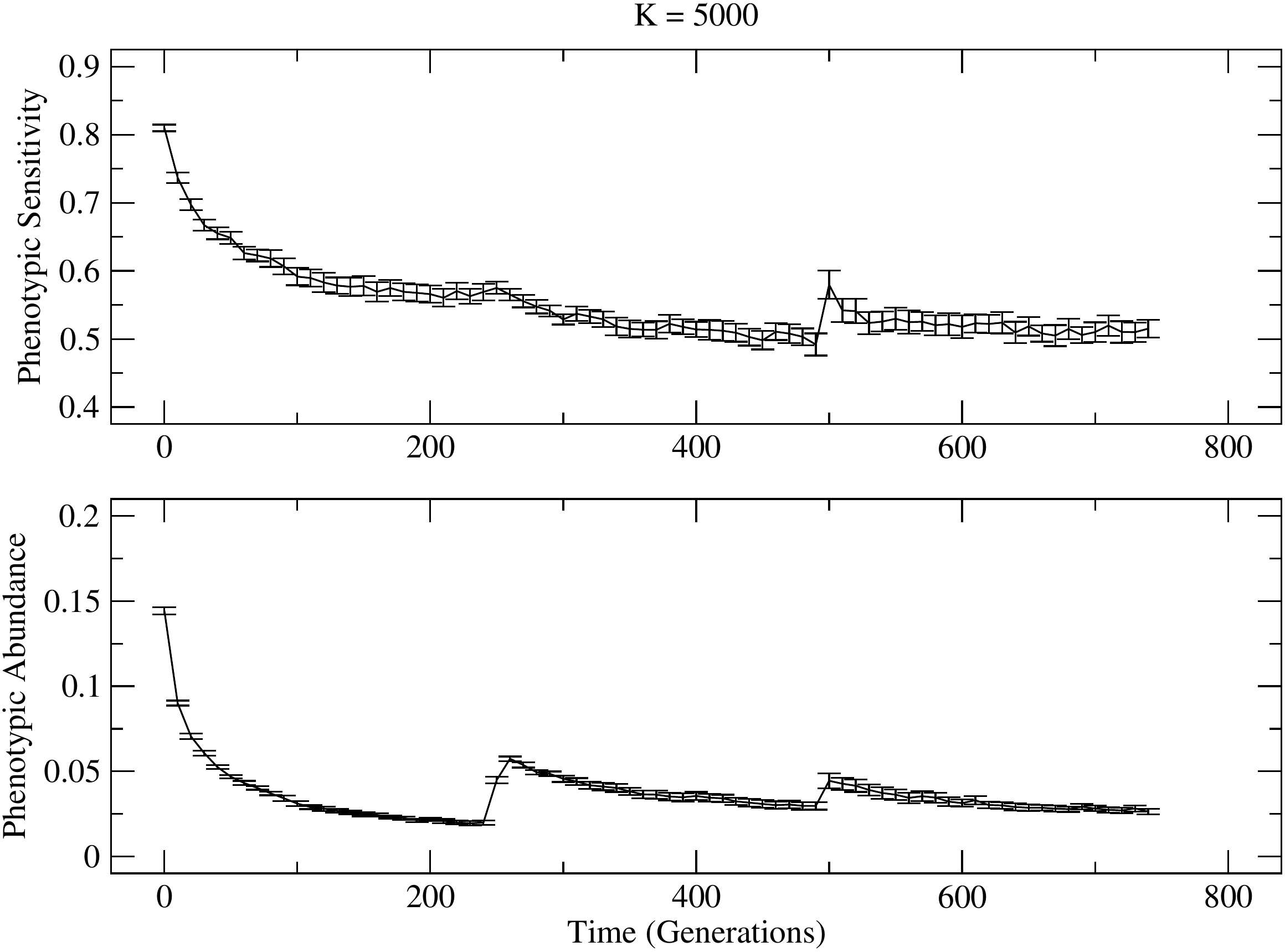}
\end{center}
\caption{Simulations of the evolution of robustness and phenotypic diversity in a model of gene regulatory networks (see text for details). The upper plot corresponds to the phenotypic sensitivity and the lower plot to the phenotypic abundance (see caption of Fig. \ref{fig:gnr-nosil}). In these simulations we have introduced a gene silencing event (see main text for details) at generation $t=250$, which is released at $t=500$. Parameter values: $N_G=10$, $\mu=K^{-1}=0.0002$, $r_m=1$  \label{fig:gnr-sil}}
\end{figure}

We now repeat the same simulatons with the difference that at a given point in time we randomly select one of the nodes, say node $i$, and introduce a gene silencing event, i.e. $g_i=-1$ which we maintain fixed as the rest of the network evolves according to Eq. (\ref{eq:gnr1}). The results are shown in Fig. \ref{fig:gnr-sil}. We observe that the induced mutation induces both an increase in the number of robust phenotypes and their robustness. These findings are consistent with experimental observations concerning liberation of cryptic genetic variation by inactivation of an evolutionary capacitor \cite{rutherford1998} as well as with recent modelling results \cite{siegal2002,bergman2003,levy2008}. 

Note that we do not explicitly deal with oscillatory asymptotic states of the gene regulatory network. In the present case, however, this is unimportant for the following reason. If from a fixed point a mutation generates an oscillation, this corresponds to a different phenotype and the phenotype switch will be correctly counted as such. If, previous to the mutation, the gene regulatory network is in an oscillatory state and the mutation does not produce a change in phenotype but simply, say, a phase shift, which does not have any bearing on the phenotype exhibited by the network, this will be counted as change in phenotype. This, however, makes our point on the increase in robustness by gene silencing even stronger, as what we are actually measuring is an lower bound of the phenotypic robustness.

Now that we have established how gene silencing affects our model of gene regulatory networks, we move on to the analysis of effect of mutations on the dynamics at the level of cell populations. In particular, we are interested in the competition between a resident (normal) population of cells and a population initiated by one cell that has undergone a gene silencing event. We investigate under which conditions the mutant will take over the resident. We start by introducing the main results of the mathematical formalism that we use to analyse this situation.  
 
\subsection*{Results for multi-type branching process with resource limitation.} In order to advance further our discussion, we focus on a multi-type branching process with resource limtation. In the context of this paper, the different types in which the population divides correspond to the robust phenotypes. We further assume that all the (pheno)types yield offspring with the same probability, namely that of an individual in our original population model. For the sake of example, we will consider two different populations. The incumbent population is defined by an \emph{strategy} whereby the population splits in $d_I$ types. The average number of offspring to be produced by a given indvidual is given by:

\begin{equation}\label{eq:6}
{\sf A}=2e^{-\mu N}\left(\begin{array}{cccc}1-\nu & \frac{\nu}{d_I-1} & \cdots & \frac{\nu}{d_I-1}\\\frac{\nu}{d_I-1} & 1-\nu & \cdots & \frac{\nu}{d_I-1}\\\frac{\nu}{d_I-1} & \cdots & \cdots & \frac{\nu}{d_I-1}\\\frac{\nu}{d_I-1} & \cdots & \cdots &1-\nu  \end{array}\right)
\end{equation}

\noindent where $\nu_{ij}$ is the mutation probability per generation from type $i$ to type $j$. For simplicity we will assume $\nu_{ij}=\nu$ $\forall i,j$. Here $N=N_I+N_M$, i.e. the total population (both resident and mutant). In all the simulations shown in this paper the initial mutant population is $N_M=1$.

The mutant population adopts a strategy with a larger variety of available types ($d_M\geq d_I$ different types) but which may have an increased robustness (smaller type mutation probability). Otherwise, the proliferation probability is the same as for the incumbent population. Thus:

\begin{equation}\label{eq:7}
{\sf B}=2e^{-\mu N}\left(\begin{array}{cccc}e^{\varphi}(1-\nu+\rho) & e^{\varphi}\frac{\nu-\rho}{d_M-1} & \cdots & e^{\varphi}\frac{\nu-\rho}{d_M-1}\\\frac{\nu-\rho}{d_M-1} & 1-\nu+\rho & \cdots & \frac{\nu-\rho}{d_M-1}\\\frac{\nu-\rho}{d_M-1} & \cdots & \cdots & \frac{\nu-\rho}{d_M-1}\\\frac{\nu-\rho}{d_M-1} & \cdots & \cdots &1-\nu+\rho  \end{array}\right)
\end{equation}

\noindent In Eq. (\ref{eq:7}), $\rho$ is a measure of the increase in resilience of the corresponding phenotype. The parameter $\phi$ corresponds to an increment in the growth rate of one of the phenotypes, which can be either positive or negative. The remaining phenotypes produce offspring at the same rate as the phenotypes of the incumbent population.

With relation to  the results of Maley et al. on the Barrett's esophagus \cite{maley2006}, the competition between the two populations in our model could represent the initial emergence of the neoplasm (mutant population) which overtakes the normal tissue (incumbent population). 

\begin{figure}
\begin{center}
\includegraphics[scale=0.5]{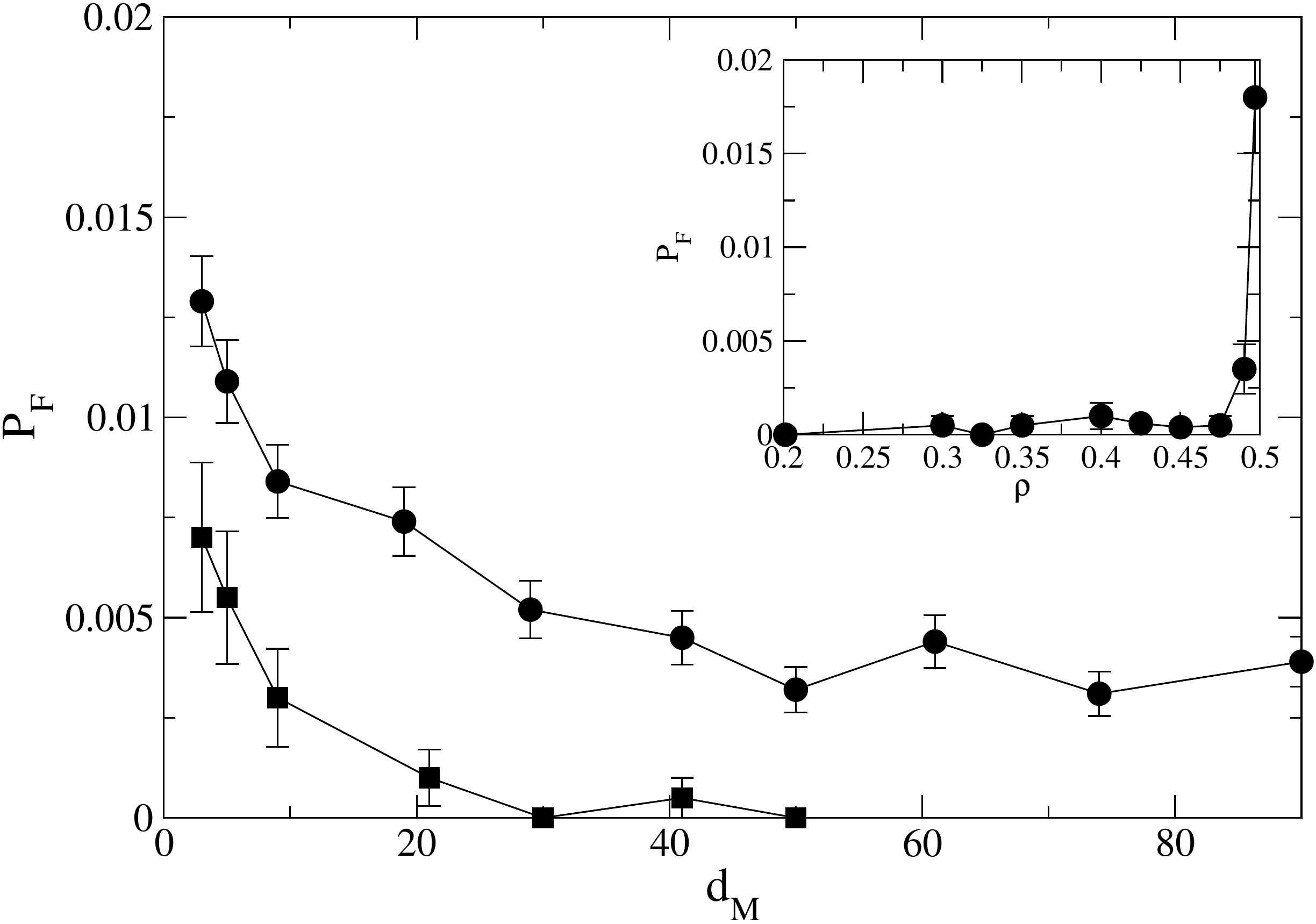}
\end{center}
\caption{This figure shows the fixation probability, $P_F$, as a function of $d_M$ corresponding to $\phi=0.01$ for $\mu=0.0001$ with $\rho=0$ (squares) and $\rho=0.49$ (circles) and $\mu=0.001$ (squares) with $\delta=0$. The inset shows the fixation probability for $\varphi=0.01$ and $\mu=0.0001$ as a function of the phenotypic resilience, $\rho$. \label{fix-prob}}
\end{figure}

\begin{figure}
\begin{center}
\includegraphics[scale=0.5]{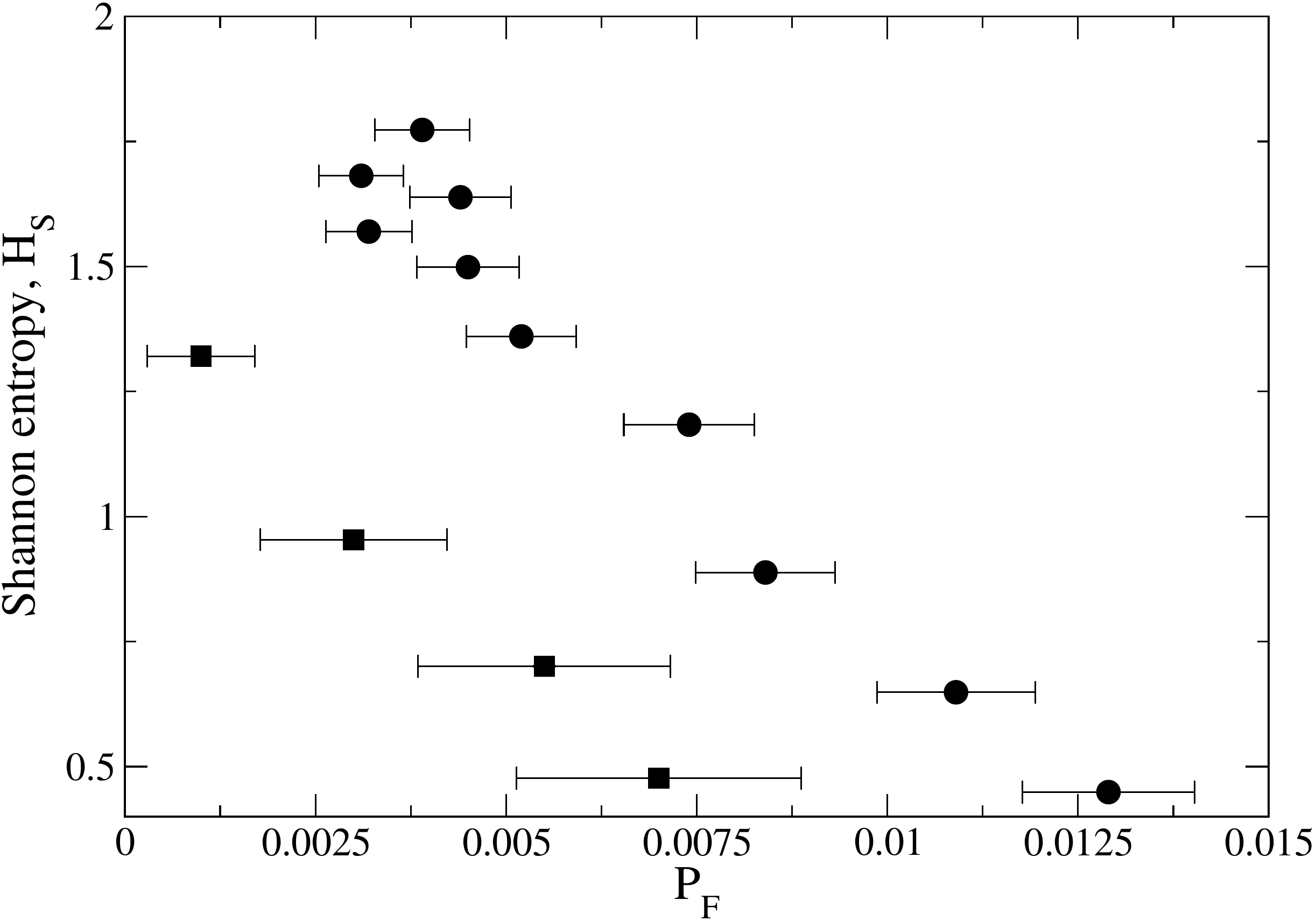}
\end{center}
\caption{This plot shows the negative correlation between the average Shannon entropy of the mutant population and the corresponding fixation probability. Circles correspond to simulation results for $\rho=0$ and squares to $\rho=0.49$.\label{shannon-entropy}}
\end{figure}

There are a number of results that we can obtain from numerical simulation of the corresponding branching process (see Materials \& Methods for details). First, increased diversity ($d_M> d_I$) without increased resilience results in loss of the ability of the mutant to invade the resident population, this is because  $P_F$ decreases when $d_M$ increases (see Fig. \ref{fix-prob}). In other words, increasing the phenotypic diversity induces an error catastrophe-like behaviour. Furthermore, increasing resilience (i.e. increasing $\rho$) rescues the more diverse mutant populations from the error catastrophe, so that they can now invade populations with less diversity, as shown in Fig. \ref{fix-prob}. Both these results can be explained in terms of the evolutionary formalism \cite{alarcon2009}. We can also observe (Fig. \ref{shannon-entropy}) that $H_S$ and $P_F$ are negatively correlated: the larger the Shannon entropy of the mutant population the less likely fixation is. In other words, high-diversity mutant populations are less likely to invade and reach fixation.

We consider now the stability and resilience to invasion of the mutants that have achieved fixation. Within the context of the problem we are dealing with, namely, the transition from neoplastic lesions to fully malignant tumours, this transition requires further mutations that perturbs the neoplastic cells and drives them into the malignant state. We model this type of mutations as perturbations in the parameters that determine the population dynamics. In particular, we consider that a second phenotype acquires a slight advantage in its proliferation probability quatified by a factor $e^{\varphi}$. We assume that perturbations in the resilience of the phenotypes is represented by an additive term $\Delta\rho$. The mean-field dynamics (see Appendix) of this second mutant population is thus described by the matrix:

\begin{equation}\label{eq:8}
\bar{{\sf B}}=2e^{-\mu N}\left(\begin{array}{cccc}e^{\varphi}(1-\nu+\rho+\Delta\rho) & e^{\varphi}\frac{\nu-\rho-\Delta\rho}{d_M-1} & \cdots & e^{\varphi}\frac{\nu-\rho-\Delta\rho}{d_M-1}\\e^{\varphi}\frac{\nu-\rho-\Delta\rho}{d_M-1} & e^{\varphi}(1-\nu+\rho+\Delta\rho) & \cdots & e^{\varphi}\frac{\nu-\rho-\Delta\rho}{d_M-1}\\\frac{\nu-\rho-\Delta\rho}{d_M-1} & \cdots & \cdots & \frac{\nu-\rho+\Delta\rho}{d_M-1}\\\frac{\nu-\rho-\Delta\rho}{d_M-1} & \cdots & \cdots &1-\nu+\rho-\Delta\rho  \end{array}\right)
\end{equation}  
 
\begin{figure}
\begin{center}
\includegraphics[scale=0.5]{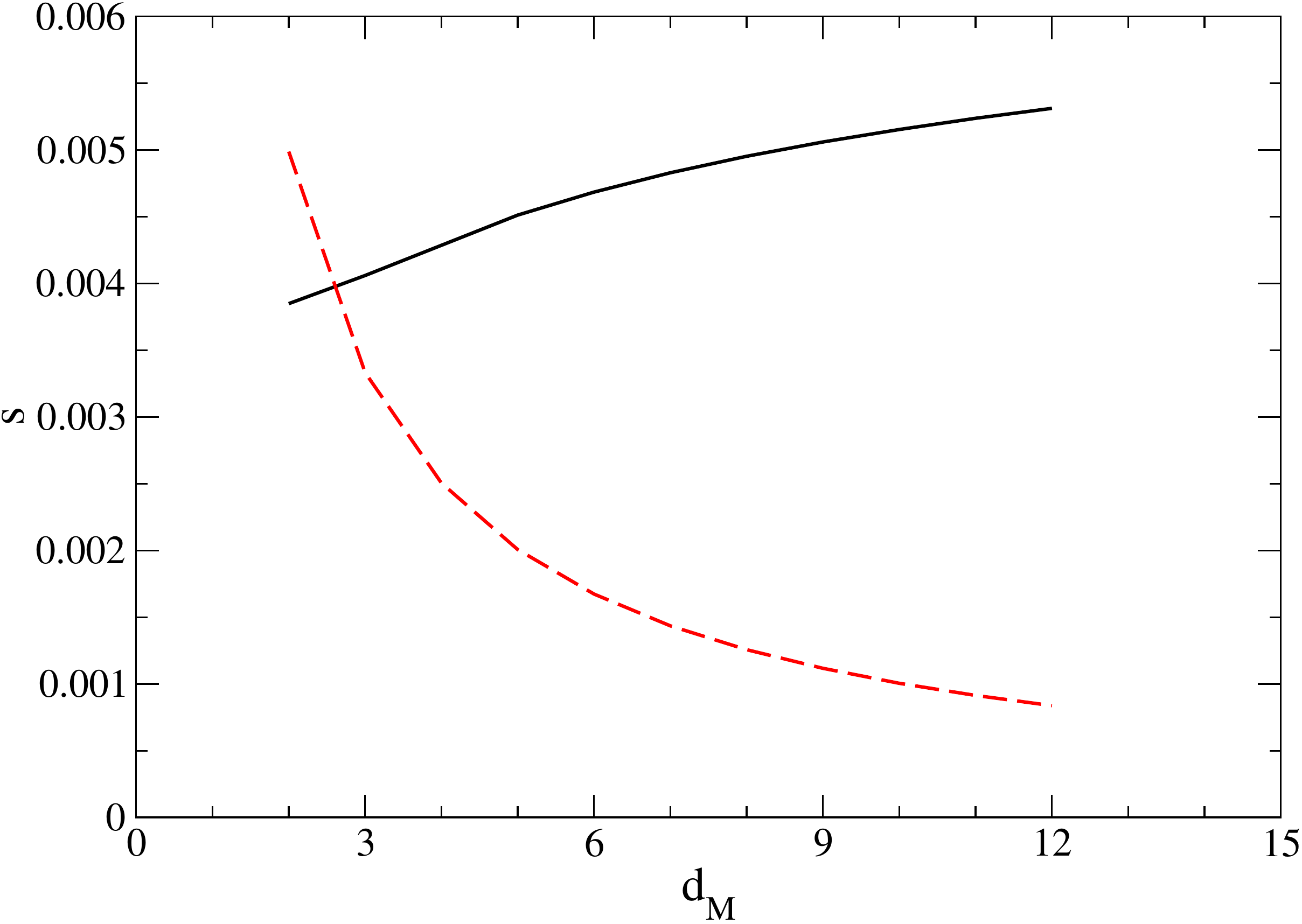}
\end{center}
\caption{This plot shows the invasion fitness, $s$, as a function of $d_M$. The red dashed line and the black solid line correspond to $\rho=0$ and $\rho=0.49$, respectively. $\Delta\rho=0.0075$ in either case. \label{fitness}}
\end{figure}

\begin{figure}
\begin{center}
\includegraphics[scale=0.5]{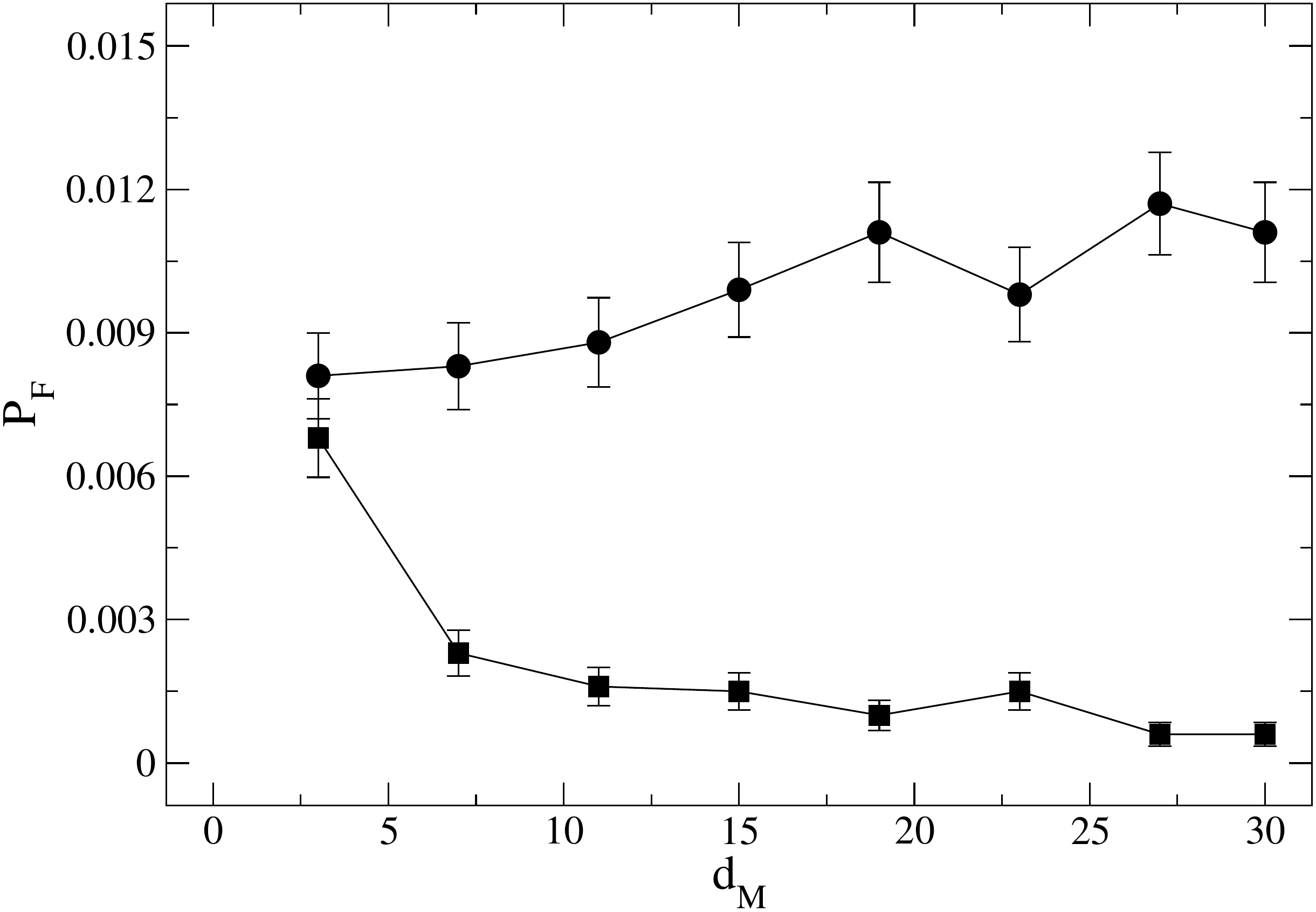}
\end{center}
\caption{This plot shows the fixation probability of a population described by $\bar{{\sf B}}$ over a ${\sf B}$-population, $s$, as a function of $d_M$. Squares correspond to $\rho=0$ and circles to $\rho=0.49$, respectively. $\Delta\rho=0.0075$ in either case. \label{second-invasion}}
\end{figure}

We consider now essentially the same problem as above, namely, the likelyhood of a population whose (deterministic) dynamics is described by $\bar{{\sf B}}$ to take over a resident population whose dynamics is given by ${\sf B}$ (Eq. (\ref{eq:7})). Using the methods described at length in our previous work \cite{alarcon2009}, the relative fitness $s$ (see Materials \& Methods for the precise definition of this quantity) of $\bar{{\sf B}}$ with respect to ${\sf B}$, which characterises the fixation probability of the former when competing with the latter, can be computed. The results are shown in Fig. \ref{fitness}. We observe that in the case in which the original mutation (that is, the one giving rise to the neoplastic lesion) increases the diversity of the population, i.e. $d_M$, without a concomitant increase in the phenotypic resilience, i.e. $\rho=0$ (see Fig. \ref{fitness}, red dashed line), the ability of the neoplasm for further evolution decreases with increasing $d_M$. In other words, the robustness of the neoplastic lesion increases with $d_M$, as the probability of it being invaded is a monotonically decreasing function of $d_M$. If, on the contrary, the original mutation, in addition to increasing the number of cellular types within the population, also increases their resilience, i.e. $\rho>0$, the robustness of the resulting population decreases with increasing $d_M$, as the corresponding relative fitness, and therefore the fixation probability, is a monotinically increasing function of $d_M$ (see Fig. \ref{fitness}, black solid line). Numerical results corresponding to this situation are presented in Fig. \ref{second-invasion} and show agreement with the theoretical predictions of the evolutionary formalism.

\section*{Discussion}

In this paper we have analysed the effects of gene silencing by means of gene silencing on a model of gene regulatory networks and we have studied how it affects phenotypic diversity and robustness. We have then investigated how these changes can modify the ability of a resident population to withstand invasion by mutants, and, last, we have discussed how our general results may shed some light on recent results regarding the evolution from neoplastic lesions to fully-malignant tumours.

We have shown that, within our model, gene silencing leads to both an increase in phenotypic diversity and also enhances the robustness, i.e. in the populations ability to sustain gene mutations. Furthermore, we have done this in a completely neutral model with no viability conditions \cite{siegal2002}. Our neutral model allows us to compute a lower bound on the robustness. 

We have then proceeded to analyse the consequences of the above observations for the competition between resident and mutant populations and on the ability of the mutant population to further evolve into more malignant variants. We have done this within the framework of multitype branching processe with resource limitation and by applying the evolutionary formalism. We have shown that increased diversity, as measured by the corresponding Shannon entropy, correlates with decreased fixation likelyhood for the initial invasion (i.e. the emergence of the neoplasm, e.g. Barrett's oesophagus). If, in addtion to an increased number of phenotypes, there is a concomitant increase in the phenotypic resilience, i.e. $\rho\neq 0$, the likelyhood of invasion increases. However, even the latter case where invasion is more likely, the same general pattern of anti-correlation between Shannon entropy and fixation probability remains (see Figs. \ref{fix-prob} and \ref{shannon-entropy}). Furthermore, we have shown that both increased phenotypic resilience and increased diversity correlate with lower robustness of the population and therefore they are more likely to be further invaded. 

These general results on population dynamics can help us to make sense of some recent data regarding the progression of the Barrett's oesophagus into a fully malignant throat cancer \cite{maley2006}. First, less than 5\% of the cases of Barrett's oesophagus actually evolved into cancer. This fact can be understood in terms of our results: lesions with increased diversity, which are the more likely ones to evolve into a tumour, are going to be less frequent, as their fixation probability is smaller. Furthermore, we predict that the fact that more diverse lesions are more likely to evolve into cancer implies that the event leading to the pre-malignant lession must also increase the phenotypic resilience, as that induces loss of robustness in more diverse populations.

\section*{Materials and Methods}

\paragraph*{Dynamics of the population of networks} In order to analyse the robustness properties of the corresponding phenotypes, $\phi$, which we define as the probability that a mutation of the first type according to the description in the paragraph above induces a phenotype switch (defined as a change in sign of at least one of the components of the vector $\phi$), we carry out a numerical experiment defined by the following algorithm:

\begin{enumerate}
\item An initial population of $N_0$ ${\sf W}$-matrices, i.e.  directed networks, are randomly generated as described above. Each of these matrices is assumed to define a particular individual within the population.
\item The gene regulatory network dynamics described above is iterated for a number of time steps (fixed to a much larger value than typically needed to reach a steady state) for each of the $N_0$ individuals. The vector $(g_1,\dots,g_{N_G})$ at the end of  iteration round is taken to be the phenotype corresponding to each individual within the population.
\item The population is next binned according to the corresponding phenotype, with $N_{\phi_j}$ being the number of individuals with a particular phenotype $\phi_j=\{g_i,\;i=1,\dots,N_G\}$ 
\item Each individual produces two offspring with probability $p_o=\mbox{e}^{-\mu N(t)}$, where $\mu$ is the carrying capacity
\item Upon division, each regulatory interaction is mutated ($w_{ij}\rightarrow -w_{ij}$) with probability $p_{m}=r_m/E$, where $r_m$ is the mutation probability per generation per edge and $E$ is the number of non-zero entries in the matrix $w_{ij}$
\item If no mutation occurs, two new individuals are incorporated in the next generation with the same $w_{ij}$ and $\phi_i$
\item If mutations happen, steps 1 to 3 are repeated and new individuals are created accordingly
\end{enumerate}

\paragraph*{Summary of the evolutionary formalism.} Here we present a brief summary of the main results and formulae of the evolutionary formalism. For a full account of the details we refer the reader to \cite{arnold1994,demetrius2004,demetrius2009}. For specific results concerning the particular model we are analysing here, see \cite{alarcon2009}. 

Consider a population with $d_I$ different types whose dynamics is given by iteration according to $u(t+1)={\sf A}(t)u(t)$, where $u(t)=(n_1(t),\dots,n_{d_I}(t))^T$ is the vector consisting of  the population sizes of each of the types and ${\sf A}=(a_{ij})$ is the average number of offspring of type $i$ produced per generation by individuals of type $j$. Note that, in general, the entries of this matrix depend on the total population $N_I=\sum_{i=1}^{d_I}n_i$. We will consider the long-time behaviour of the population, i.e. once it has settled in a steady state. We will further assume that ${\sf A}$ is an irreducible matrix. 

The long time behaviour of the population is determined by the dominant eigenvalue of the matrix ${\sf A}$, which will denoted by $\lambda_0$. Let ${\sf u}$ and ${\sf v}$ be the corresponding left and right eigenvectors, respectively: ${\sf A}{\sf u}=\lambda_0{\sf u}$ and ${\sf v}{\sf A}=\lambda_0{\sf v}$. We can now define an associated Markov matrix, ${\sf P}$, such that:

\begin{equation}\label{eq:1}
p_{ij}=\frac{a_{ji}v_j}{\lambda_0v_i} 
\end{equation}

One of the main results consists in a variational principle, which states that the following relation between the growth rate $r=\log\lambda_0$, the evolutionary entropy, $H_E$, and the so-called proliferative potential, $F$, \cite{arnold1994}:

\begin{eqnarray}\label{eq:2}
\nonumber && r=H_E+F \\
\nonumber && H_E=\sum_{i,j=1}^{d_I}\pi_ip_{ij}\log p_{ij} \\
&& F=\sum_{i,j=1}^{d_I}\pi_ip_{ij}\log a_{ij}
\end{eqnarray}

\noindent where $\pi_i$ is the stationary distribution corresponding to the Markov process defined by ${\sf P}$. The requirement that ${\sf A}$ should be irreducible ensures the uniqueness of this distribution.

A number of interesting results follow from Eq. (\ref{eq:2}) and the corresponding variational principle, for example, a proof concerning large deviations  states that the rate of relaxation to the steady state after a perturbation of the entries of the matrix ${\sf A}$, specifically, a perturbation of the type ${\sf A}(\delta)=(a_{ij}^{1+\delta})$, is positively correlated with the corresponding change in the evolutionary entropy $H_E(\delta)-H_E(\delta=0)$, thus proving that this quantity is a measure of robustness of the population \cite{demetrius2004}. 

However, we are more interested in a (related) result concerning the evolutionary stability of an incumbent population against an invading  mutant population when both compete for a common limited resource. By means of a diffusion approximation, Demetrius et al. \cite{demetrius2009} have showed that the fixation probability of the mutant, $P_F$ is given by:

\begin{equation}\label{eq:3}
P_F(y)=\frac{1-\left(1-\frac{\Delta\sigma^2}{\sigma^2_M}y\right)^{\frac{2\langle N\rangle s}{\Delta\sigma^2}+1}}{1-\left(1-\frac{\Delta\sigma^2}{\sigma^2_M}\right)^{\frac{2\langle N\rangle s}{\Delta\sigma^2}+1}}
\end{equation}

\noindent where the total population is assumed to be constant, $y$ is the initial proportion of mutants, and $s$ is defined by:

\begin{equation}\label{eq:4}
s=\Delta r-\frac{\Delta\sigma^2}{\langle N\rangle}
\end{equation}

\noindent where $\langle N\rangle$ is the average stationary population. In the case of the branching process we consider in this paper, $\langle N\rangle$ is given by $2e^{-\mu\langle N\rangle}=1$, i.e. $\mu\langle N\rangle=\log 2$. The quantity $\Delta r$ is defined as $\Delta r\equiv r_M-r_I$, where the subindices $I$ and $M$ denote the corresponding quantities for the incumbent and mutant populations, respectively. The quantity $\Delta \sigma^2$ is defined as $\Delta\sigma^2=\sigma^2_M-\sigma^2_I$ where \cite{alarcon2009}:

\begin{equation}\label{eq:5}
\sigma^2=-\left.\frac{dH_E(\delta)}{d\delta}\right\vert_{\delta=0}
\end{equation}

The behaviour of $P_F$ can be analysed in terms of the quantity $s$. If $s>0$, $P_F(y)$ is convex and $P_F(y)>y$. If, on the contrary, $s<0$ then $P_F(y)$ is concave and $P_F(y)<y$. This means that if $s<0$ it is very likely that the mutant gets extinct whereas if $s>0$ it is very likely that the mutant is fixed, with the likelihood of invasion increasing as $s$ increases. Moreover, the larger the value of $s$ the more likely fixation is. Therefore $s$ can be taken as a measure of fitness. 

\paragraph*{Definition of the muitype branching processes}

To precisely define the branching processes involved in our models of population dynamics, we use the corresponding generating function formalism \cite{kimmel2002} where a generating function for the probability distribution of the per individual offspring number is prescribed. In this article we consider three different populations (corresponding to the matrices ${\sf A}$, ${\sf B}$, and $\bar{{\sf B}}$). The set of generating functions corresponding to the population associated to the matrix ${\sf A}$ is:

\begin{equation}\label{eq:app1}
g_i(\vec{z})=(1-e^{-\mu N})+(1-\nu)e^{-\mu N}z_i^2+\sum_{j\neq i}\frac{\nu}{d_I-1}e^{-\mu N}z_j^2
\end{equation}

\noindent where $i=1,\dots,d_I$ and $\vec{z}$ is a $d_I$-dimensional vector with $0\leq z\leq 1$.

Similarly the generating function for the populations associated to ${\sf B}$ and $\bar{{\sf B}}$ are, respectively:

\begin{eqnarray}\label{eq:app2}
\nonumber g_1(\vec{z})= && (1-e^{-\mu N+\varphi})+(1-(\nu-\rho))e^{-\mu N+\varphi}z_1^2\\
\nonumber && +\sum_{j\neq 1}\frac{\nu-\rho}{d_M-1}e^{-\mu N+\varphi}z_j^2\\
\nonumber g_i(\vec{z})= && (1-e^{-\mu N})+(1-(\nu-\rho))e^{-\mu N}z_i^2+\\
&& \sum_{j\neq i}\frac{\nu-\rho}{d_M-1}e^{-\mu N}z_j^2\mbox{ for }i\geq 2
\end{eqnarray}

\noindent and  

\begin{eqnarray}\label{eq:app3}
\nonumber g_i(\vec{z})= && (1-e^{-\mu N+\varphi})+(1-(\nu-\rho-\Delta\rho))e^{-\mu N+\varphi}z_i^2\\
\nonumber && +\sum_{j\neq i}\frac{\nu-\rho-\Delta\rho}{d_M-1}e^{-\mu N+\varphi}z_j^2\mbox{ for }i\leq 2\\
\nonumber g_i(\vec{z})=&&(1-e^{-\mu N})+(1-(\nu-\rho-\Delta\rho))e^{-\mu N}z_i^2\\
&& +\sum_{j\neq i}\frac{\nu-\rho-\Delta\rho}{d_M-1}e^{-\mu N}z_j^2\mbox{ for }i > 2
\end{eqnarray}

\noindent For both Eqs. (\ref{eq:app2}) and (\ref{eq:app3}) where $i=1,\dots,d_M$ and $\vec{z}$ is now a $d_M$-dimensional vector with $0\leq z\leq 1$.

In our simulations, at each generation we go through all individuals and kill them with a probability given by the 0th-order term in the generating functions Eqs. (\ref{eq:app1}), (\ref{eq:app2}) and (\ref{eq:app3}), thus passing zero individuals on to the next generation. Otherwise, two individuals are passed on to the next generation of the same type, say type $i$, with probability given by the coefficient of the $y_i^2$- or $z_i^2$-terms or of a different type, say type $j$, with probability equal to the coefficent of the $y_i^2$- or $z_i^2$-terms.

Note that the entries matrices ${\sf A}$, ${\sf B}$, and $\bar{{\sf B}}$ correspond to the branching ratios $m_{ij}=\partial_jg_i({\bf 1})$ as defined in Eqs. (\ref{eq:app1}), (\ref{eq:app2}) and (\ref{eq:app3}), respectively, and therefore they determine the mean-field dynamics. 

\section*{Acknowledgments}
TA and HJJ gratefully acknowledge the EPSRC for funding under grant EP/D051223.


\end{document}